\documentclass[12pt,a4paper]{article}

\usepackage[utf8]{inputenc}
\usepackage{amsfonts}
\usepackage{amsmath}
\usepackage{graphicx}
\usepackage[T1]{fontenc}
\usepackage{amssymb}
\usepackage{comment}

\usepackage{natbib}
\usepackage{endfloat} 
\usepackage{authblk} 
\usepackage{setspace}

\title{\small \textbf{Title:} On the importance of being structured: instantaneous coalescence rates and a re-evaluation of human evolution}

\author[1,$\S$]{\small \textbf{Authors:} Mazet Olivier} 	
\author[1, $\S$]{Rodríguez Willy}
\author[1]{Grusea Simona}
\author[2,3]{Boitard Simon}
\author[4,5,6]{Chikhi Lounès}				

\affil[1]{\footnotesize Université de Toulouse, Institut National des Sciences Appliquées, Institut de Mathématiques de Toulouse, F-31077 Toulouse, France}
\affil[2]{\footnotesize UMR7205 Institut de Systématique, Evolution et Biodiversité, Ecole Pratique des Hautes Etudes \& Muséum National d'Histoire Naturelle \& CNRS \& Université Pierre et Marie Curie, Paris, France}
\affil[3]{\footnotesize UMR1313 Génétique Animale et Biologie Intégrative, Institut National de la Recherche Agronomique \& AgroParisTech, Jouy-en-Josas, France}
\affil[4]{\footnotesize CNRS, Université Paul Sabatier, ENFA, UMR 5174 EDB (Laboratoire Évolution \& Diversité Biologique), F-31062 Toulouse, France}
\affil[5]{\footnotesize Université de Toulouse, UPS, EDB, F-31062 Toulouse, France}
\affil[6]{\footnotesize Instituto Gulbenkian de Ciência, P-2780-156 Oeiras, Portugal}

\affil[$\S$] {These authors contributed equally to this work.}

\date{
  \small \textbf{Corresponding author:} Lounès Chikhi / Instituto Gulbenkian de Ciência / P-2780-156 Oeiras, Portugal / +351 214464671 / lounes.chikhi@univ-tlse3.fr\\ 
  \textbf{Running title: } population structure and demographic inference\\
  \textbf{Word count:} $7191$
 }

\begin{document}
\maketitle 

\doublespacing

\newpage
\section*{Abstract}
Most species are structured and influenced by processes that either increased or reduced gene flow between populations. However, most population genetic inference methods ignore population structure and reconstruct a history characterized by population size changes under the assumption that species behave as panmictic units. This is potentially problematic since population structure can generate spurious signals of population size change. Moreover, when the model assumed for demographic inference is misspecified,  genomic data will likely increase the precision of misleading if not meaningless parameters. In a context of model uncertainty (panmixia \textit{versus} structure) genomic data may thus not necessarily lead to improved statistical inference.

We consider two haploid genomes and develop a theory which explains why any demographic model (with or without population size changes) will necessarily be interpreted as a series of changes in population size by inference methods ignoring structure. We introduce a new parameter, the IICR (inverse instantaneous coalescence rate), and show that it is equivalent to a population size only in panmictic models, and mostly misleading for structured models. We argue that this general issue affects all population genetics methods ignoring population structure. We take the PSMC method as an example and show that it infers population size changes that never took place. We apply our approach to human genomic data and find a reduction in gene flow at the start of the Pleistocene, a major increase throughout the Middle-Pleistocene, and an abrupt disconnection preceding the emergence of modern humans.

\noindent \textbf{KeyWords:} coalescence rate, failure rate, population size change, demographic history.

%
%

\newpage
\section{Introduction}

Most species are structured, and do not behave as panmictic populations \citep{wakeley1999nonequilibrium,harpending2000genetic,goldstein2002human,charlesworth2003structure,harding2004structured}. They have been influenced by habitat fragmentation, expansion or reconnection events that either increased or reduced the amount of gene flow between local populations, as a result of climatic or anthropogenic events \citep{goossens2006genetic,quemere2012genetic}. While genomic data offer the possibility to reconstruct with increasing precision major events in that complex history \citep{gutenkunst2009inferring,li2011inference,Sheehan22042013}, it is computationally very difficult to account for population structure. As a consequence, many inferential methods tend to ignore population structure \citep{li2011inference,Sheehan22042013,schiffels2013msmc}. This is potentially problematic because an increasing number of studies have shown that population structure generates spurious signals of changes in population size, even when populations were stationary \citep{wakeley1999nonequilibrium,nielsen2009statistical,chikhi2010confounding,peter2010distinguishing,heller2013confounding,paz2013demographic}. 
Here, we provide a simple theoretical framework which explains why any inferential method ignoring population structure will always infer population size changes as soon as populations are actually structured. In other words, this theory explains why any real demographic history, with or without structure, will necessarily and optimally be interpreted as a series of changes in population size by methods ignoring population structure.

We consider the case of two haploid genomes and we study $T_2$, the coalescence time for a sample of size two (\textit{i.e.} the time to the common ancestor of two randomly sampled sequences   \citep{herbots1994stochastic,Mazet201546}). We predict the history that any population genetics methods ignoring structure will try to reconstruct. We introduce a new parameter, which we call the IICR (inverse instantaneous coalescence rate). The IICR is equivalent to a population size only in panmictic models. For models incorporating population structure the IICR exhibits a temporal trajectory that can be strongly disconnected from the real demographic history (\textit{i.e.} identifying a decrease when the population size was actually constant or increasing). 

We apply our approach to simulated data and use the PSMC of Li and Durbin \citep{li2011inference} as a reference method because it allows to reconstruct the history of a population or species from one single diploid genome. Also, this method has been applied to a wide array of vertebrate species including reptiles \citep{green2014crocodile}, birds \citep{zhan2013psmcperegrine,hung2014pigeon} and mammals such as primates \citep{prado2013psmcgreatapes,zhou2014snubgenome}, pigs \citep{groenen2012pig} and pandas \citep{zhao2013panda} and its outcome has been interpreted in terms of population size changes.

We then apply our approach to human data and show that an alternative model involving a minimum of three changes in migration rates can explain the PSMC results obtained by Li and Durbin. The history that we find suggests a reduction in gene flow around 2.55 MY ago, at the start of the Pleistocene (2.58 MY), when the \textit{Homo} genus differentiated from australopithecines. This is followed by a major increase in gene flow around 0.8-1.0 MY ago, just before the transition from Lower to Middle Pleistocene (MP, at 0.78 MY). The MP is then characterized by sustained gene flow until 200-230 KY ago when an abrupt decrease in gene flow coincides with the emergence of anatomically modern humans. These results represent an alternative to the population crashes and increases depicted in many population genetic studies, but it is strikingly in phase with fossil data and provides a more realistic framework as several authors have sugested \citep{goldstein2002human,harding2004structured}. Altogether we call for a major re-evaluation of what genomic data can actually tell us on the history of our species. Beyond our species we argue that genomic data should be re-interpreted as a consequence of changes in levels of connection rather than simple changes in population size.

\section{Models, Theory}

\subsection*{Coalescence time for a sample of size 2 in a model of population size change}

We consider a model of arbitrary population size change (PSC), where $N(t)$ represents the population size ($N$, in units of genes or haploid genomes) as a function of time $(t)$ scaled by the number of genes (\textit{i.e.} in units of coalescence time, corresponding to $\lfloor N(0)t\rfloor$ generations). We consider that $t=0$ is the present, and positive values represent the past. Since $N$ represents the population size in terms of haploid genomes, the number of individuals will be $N/2$ for diploid species. We can then apply the generalisation of the coalescent in populations of variable size \citep{donnelly1995coalescents,nordborg2001coalescent,tavare2004part}. 
If we denote by $\lambda(t)$ the ratio $\frac{N(t)}{N(0)}$,
 we can then compute the probability density function ($pdf$) $f^{PSC}_{T_2}(t)$ of the coalescence time $T_2$ of two genes sampled in the present-day population. Indeed, the probability that two genes will coalesce at a time greater than $t$ is
 
\begin{equation}
\mathbb{P}(T_2>t)=e^{-\int_0^t \frac{1}{\lambda(x)}\, dx}
\end{equation}
Given that  

\begin{equation}
f^{PSC}_{T_2}(t) =(1-\mathbb{P}(T_2>t))'
\end{equation}

\noindent we can write the $pdf$ as

\begin{equation}
\label{Eq:PSCT2density}
f^{PSC}_{T_2}(t) =(1-e^{-\int_0^t \frac{1}{\lambda(x)}\, dx})'=\frac{1}{\lambda(t)}e^{-\int_0^t \frac{1}{\lambda(x)}\, dx}
\end{equation}

Consequently, if we know the $pdf$ of the coalescence time $T_2$, the corresponding population size change function $\lambda(t)$ can be computed as:

\begin{equation}
\label{Eq:lambdapsc}
\lambda(t)=\frac{\mathbb{P}(T_2>t)}{f^{PSC}_{T_2}(t)}
\end{equation}

This means that if we only had access to a finite set of $T_2$ values we could in theory infer the history $\lambda(t)$ by simply computing this ratio. In the case of a model of population size change this computation is by definition giving us the actual history of population size change. We show below how this ratio can be computed for \textit{any} demographic scenario for which $T_2$ distributions can be derived or simulated.

\subsection*{Instantaneous coalescence rate (ICR) for a sample of size 2}

If we consider now the coalescence time of two genes sampled in a population under an arbitrary model, whichever model this may be (structured or not, with population size change or not, etc.), and if we assume that we know its $pdf$, $f_{T_2}(t)$, it is straightforward to compute the ratio $\lambda(t)$ of equation (\ref{Eq:lambdapsc})

\begin{equation}\label{Eq:lambdagen}
\lambda(t)=\frac{\mathbb{P}(T_2>t)}{f_{T_2}(t)}
\end{equation}
Let us now denote $g(t)=\mathbb{P}(T_2>t)$. We then have by definition $f_{T_2}(t)=-g'(t)$, hence

\begin{equation}
\frac{1}{\lambda(t)}=-\frac{g'(t)}{g(t)}=-\log(g(t))'
\end{equation}
from where we get, since $g(0)=1$,

\begin{equation}
g(t)=e^{\log(g(t))}=e^{-\int_0^t \frac{1}{\lambda(x)}\, dx}
\end{equation}

\noindent It therefore follows that the $pdf$ $f_{T_2}(t)=-g'(t)$ can always be written as

\begin{equation}
\label{Eq:PSCpdf}
f_{T_2}(t) =\frac{1}{\lambda(t)}e^{-\int_0^t \frac{1}{\lambda(x)}\, dx}
\end{equation}

\noindent even if the so-computed function $\lambda(t)$ has nothing to do with any population size change.

In other words, for any given model, there always exists a function $\lambda(t)$ which \textit{explains} the coalescence time distribution of this model for a sample of size two, $f_{T_2}(t)$. The $pdf$ of $T_2$ can thus always be written as a function of $\lambda(t)$ as in equation (\ref{Eq:PSCpdf}), exactly as if the model under which the data were produced was \textit{only defined by population size changes}. This function $\lambda(t)$ is a fictitious or spurious population size change function whose coalescence time $T_2$ would \textit{mimic} perfectly the demographic model.

Now, if we define $\mu(t)$ as

\begin{equation}
\label{Eq:ICR}
 \mu(t)=\frac{1}{\lambda(t)}=\frac{f_{T_2}(t)}{\mathbb{P}(T_2>t)}
\end{equation}

\noindent it should be natural to see $\mu(t)$ as an \textit{instantaneous coalescence rate} (ICR),
as it represents the probability that two lineages which have not yet coalesced at time $t$ (as expressed by the denominator), will do so in an infinitesimal amount of time starting at $t$ (as expressed in the numerator).
Another way to realize it is to note that $T_2$ can be seen as a \textit{lifetime}. 
Then, we can note that the quantity $\mu(t)=\frac{1}{\lambda(t)}$, known as the \textit{hazard function} or \textit{failure rate} in the reliability engineering community,
represents the instantaneous rate of failure of a system at time $t$ (see for instance \cite{ruegg1989processus} or \cite{klein2003survival}). 
The term \textit{instantaneous} is central and we show in the next section that it is crucial for the interpretation of structured models. 

\subsection*{Linking population structure and population size change}

We now consider a model of population structure such as the classical symmetric \textit{n-island} model \citep{wright1931evolution}, where we have a set of $n$ islands (or demes) of constant size $N$, interconnected by gene flow with a migration rate $m$, where $\frac{M}{2}=Nm$ is the number of immigrants (genes) in each island every generation. The total number of genes or haploid genomes in the whole metapopulation is $nN$ and it is therefore constant. Again, $N$ is the number of haploid genomes, and $N/2$ the number of diploid individuals. We call this model the StSI, which stands for Structured Symmetrical Island model, as in \cite{Mazet201546}.

Under this model we can write the $pdf$ for $T_2$ (see \cite{herbots1994stochastic, wilkinson1998genealogy,Mazet201546} for details and Bahlo and Griffiths \citep{bahlo2001coalescence} for related results and Charlesworth \textit{et al.} \citep{charlesworth2003structure} for an insightful review) by considering the cases when the two genes are sampled from the same ($s$) or from different ($d$) demes. 

\begin{equation}
\label{Eq:StSIT2densitys}
f^{StSI}_{T_2^s}(t)=ae^{-\alpha t}+(1-a)e^{-\beta t}
\end{equation}

\begin{equation}
\label{Eq:StSIT2densityd}
f^{StSI}_{T_2^d}(t)=ce^{-\alpha t}-ce^{-\beta t}
\end{equation}

where

\begin{equation}
a=\frac{\gamma-\alpha}{\beta-\alpha}, ~~ c=\frac{\gamma}{\beta-\alpha}
\end{equation}

and where $-\alpha$ and $-\beta$ are the roots of the polynomial 

\begin{equation}
\theta^2+\theta(1+n\gamma)+\gamma
\end{equation}

whose discriminant is $\Delta=(1+n\gamma)^2-4\gamma$, and therefore 

\begin{equation}
\label{Eq:alpha}
\alpha=\frac{1}{2}\left(1+n\gamma+\sqrt{\Delta}\right)
\end{equation}

and

\begin{equation}
\label{Eq:beta}
\beta=\frac{1}{2}\left(1+n\gamma-\sqrt{\Delta}\right)
\end{equation}

with $\gamma=\frac{M}{n-1}=\alpha\beta$.

Now let us consider a hypothetical demographic history characterized by population size changes but without any population structure. For that history to explain the data generated by a model of population structure, this hypothetical demographic history will correspond to the function $\lambda(t)$ as defined by equation \ref{Eq:lambdagen}. Thus, in the case of two haploid genomes sampled in the same deme (a most reasonable assumption for a diploid individual) we get:

\begin{equation}
\label{Eq:IICRs}
\lambda_s(t)=\frac{\mathbb{P}(T_2>t)}{f^{StSI}_{T^s_2}(t)}=\frac{\frac{a}{\alpha}e^{-\alpha t}+\frac{1-a}{\beta}e^{-\beta t}}{ae^{-\alpha t}+(1-a)e^{-\beta t}}
=\frac{(1-\beta)e^{-\alpha t}+(\alpha-1)e^{-\beta t}}{(\alpha-\gamma)e^{-\alpha t}+(\gamma-\beta)e^{-\beta t}}
\end{equation}

\begin{figure}[h]
\includegraphics[scale=0.7]{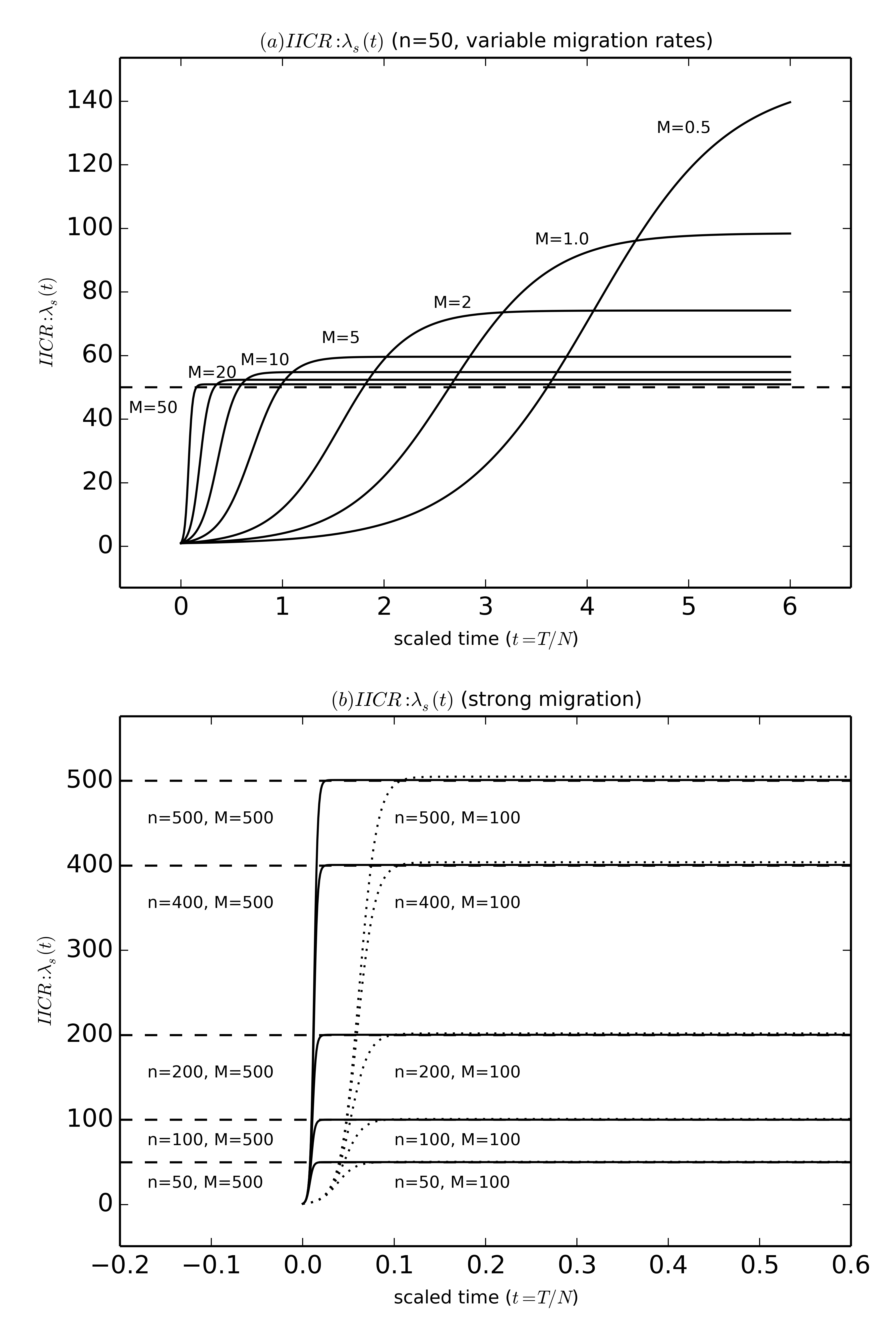}
\caption{Inferred population size changes for n-island models with constant size. This figure shows $\lambda_s(t)$ for different values of $M$, the number of migrants, and $n$, the number of islands. In panel (a) we assumed an island model with $n=50$, and varied $M$, the number of migrants between 0.5 and 50. In panel (b) we varied $n$ between $50$ and $500$ and used two large values for $M$, namely $100$ and $500$. For both panels, the $y$ axis is scaled by $N$ and the horizontal dashed lines correspond to $nN$, the total population size. In all cases, $\lambda_s(t)$ identifies a population decrease.}
\label{Fig.popsize_lambdaS}
\end{figure}

It is then trivial to compute the function $\lambda_s(t)$ for any set of parameters $n$ and $M$. Figure \ref{Fig.popsize_lambdaS} shows for instance in panel (a) the corresponding curves for $n=50$ and $M$ values between $0.5$ and $50$. As expected \citep{chikhi2010confounding,Mazet201546} we observe a (fictitious) population decrease from a large hypothetical ancestral population of size $N^h_a$ to a smaller hypothetical current population of size $N^h_c$. 
Note that $\lambda_s(t)$ is a population size ratio, which does not provide absolute values of the effective population size. In our case, it is however trivial to show that for $t$ sufficiently close to 0, we find that $\lambda_s(t)=1$ and hence it follows that $N_c^h=N$, the size of a deme. Indeed, at the time of sampling, the coalescence history for two genes sampled from the same deme is mostly dependent on the size of the local deme. Interestingly, this is true for any value of $M$.
Figure \ref{Fig.popsize_lambdaS} indicates that as $M$ becomes larger,  $\displaystyle{N_a^h=N \lim_{t\to +\infty}\lambda_s(t)}$ becomes closer to $nN$, represented by the horizontal dashed line. This is expected: when the migration rate increases the whole set of populations behaves less and less like a structured model and increasingly like a single random mating 
population of size $nN$.  Several authors have shown that under the strong migration condition, it is possible to define a coalescent effective population size towards which the structured population tends \citep{sjodin2005meaning,wakeley2009extensions}. Panel (b) shows indeed that when $M$ is very high ($M=100$ and $M=500$) the n-island model behaves as a population characterized by a constant size until the very recent past. For instance, when $M=500$, $\lambda_s(t)$ only drops at time $t=0.02$, which for $N=100$ would correspond to 2 generations ago. In other words, the strong migration assumption implicitly assumes that the bottleneck seen in our results is so recent that it can be neglected. Using the terminology introduced by \cite{wakeley1999nonequilibrium}, it assumes that the scattering phase is very short. Altogether our results provide a more general framework which allows us to easily incorporate the strong migration assumptions.

Coming back to panel (a) we also note that as $M$ decreases, the fictitious bottleneck becomes older and the ancestral population becomes larger, for a constant value of $n$, the number of islands.
We can derive the asymptotic coalescent effective size of this n-island model by computing the limit of $\lambda(t)$ when $t$ goes to infinity, and find that, since $0<\beta<\alpha$,

\begin{equation}
N_a^h=N \lim_{t \to +\infty}\lambda_s(t)=N\frac{\alpha-1}{\gamma-\beta}=\frac{N}{\beta},
\end{equation}
where we recall that $\beta$ was the smallest of the roots found above (equation \ref{Eq:beta}). By developing equation \ref{Eq:beta}, we find

\begin{equation}
\label{Eq:betafull}
\beta=\frac{1}{2}\left(1+\frac{n}{n-1}M-\sqrt{\left(1+\frac{n}{n-1}M\right)^2-\frac{4M}{n-1}}\right)
\end{equation}

Here we can see that for large values of $M$, $\lambda_s(t)$ is close to 

\begin{equation}
N^h_a=N(n+\frac{(n-1)^2}{nM})
\end{equation}

This is the nucleotide diversity effective size computed in \cite{nei1993effective} for the n-island model.

If we now perform the same analyses and computations for the case where the haploid genomes are sampled from different demes leads to the following result:

\begin{equation}
\label{Eq:IICRd}
\lambda_d(t)=\frac{\frac{1}{\alpha}e^{-\alpha t}-\frac{1}{\beta}e^{-\beta t}}{e^{-\alpha t}-e^{-\beta t}}=\frac{\beta e^{-\alpha t}-\alpha e^{-\beta t}}{\gamma e^{-\alpha t}-\gamma e^{-\beta t}}
\end{equation}

\begin{figure}[h]
\includegraphics[scale=0.7]{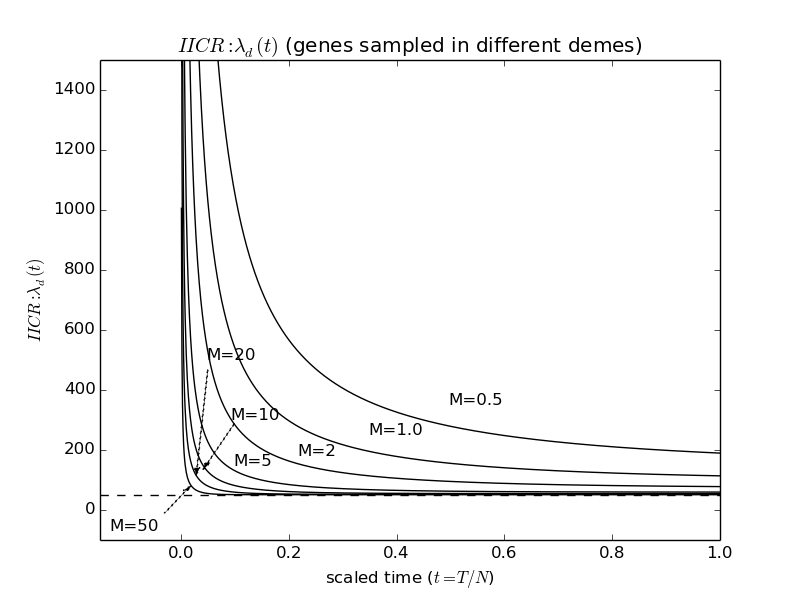}
\label{Fig.popsize_lambdaD}
\caption{Inferred population size changes for n-island models and samples from different demes. This figure shows $\lambda_d(t)$ for different values of $M$, the number of migrants. The number of islands, was assumed to be $n=50$. Samples come from different islands. In all cases, $\lambda_d(t)$ identifies a population increase.}
\end{figure}

Here the population dynamics is inverted, and we observe a fictitious population expansion. This is in agreement with several previous studies which noted that when sampling is carried out across demes the bottleneck signal either disappears or can be replaced by a population expansion signal \citep{peter2010distinguishing,chikhi2010confounding,heller2013confounding}. We note that $\displaystyle{\lim_{t\to 0}(\lambda_d(t))=+\infty}$. The two lineages being in different demes at time $t=0$, it is by definition impossible for them to coalesce in the very recent past, since a migration event has first to occur. Let us note also that $\displaystyle{\lim_{t\to \infty}(\lambda_d(t))=\frac{1}{\beta}}$ as for $\lambda_s$.

Our results, as expressed by equations (\ref{Eq:IICRs}) and (\ref{Eq:IICRd}), stress the difficulty in defining an effective size for a structured population, because a structured population has properties that a non structured population does not have. Since it behaves like a non-structured population that changes in size, there is no overwhelming reason to summarize its properties by one single number when it actually is defined either by a number of islands and a migration rate, or by a full trajectory of effective sizes. We point towards the studies of \cite{sjodin2005meaning} and \cite{wakeley2009extensions} for models and conditions under which an effective size can be defined.  What we wish to stress is that the theory presented here provides a general framework for explaining and predicting population size changes that population genetics methods will infer. Below, we illustrate how this can be applied to simple and complex structured models and we also predict the population size changes that methods ignoring structure will infer. 
Given that $\lambda(t)$ does not necessarily correspond to actual changes in $N_e$ we introduce the inverse ICR or IICR, which we will use for the rest of the manuscript instead of  $\lambda(t)$. The reason for this is that the IICR is only equivalent to an instantaneous coalescent $N_e$ in the case of models without structure. For other models, it is, in the absence of a better term, the inverse of an instantaneous coalescence rate. The IICR is of course by definition a function of time and implicitly leads us to consider a trajectory rather than a single value even for constant size models such as the StSI. 

\subsection*{Application to simulated data}

In order to illustrate how an observed distribution of $T_2$ values can be used to infer the IICR we carried out simulations under \textit{structured} and \textit{unstructured} scenarios. Data were simulated using the \textit{ms} software \citep{hudson2002ms}. For each scenario, we simulated independent values of $T_2$ and used them to estimate the IICR at various time points $t_i$, as follows:

\begin{equation}
\label{Eq:IICRti}
IICR(t_i)=\frac{1-F_{T_2}(t_i)}{f_{T_2}(t_i)}
\end{equation}

\noindent where $F_{T_2}$ is the empirical cumulative distribution function of $T_2$ and $f_{T_2}$ is an empirical approximation of its density.

\section*{Results}

\subsection*{Predicting the inferred demographic history of non structured and structured populations: illustrations by simulations}

\begin{figure}[h]
  \centering
  \includegraphics[width=\textwidth]{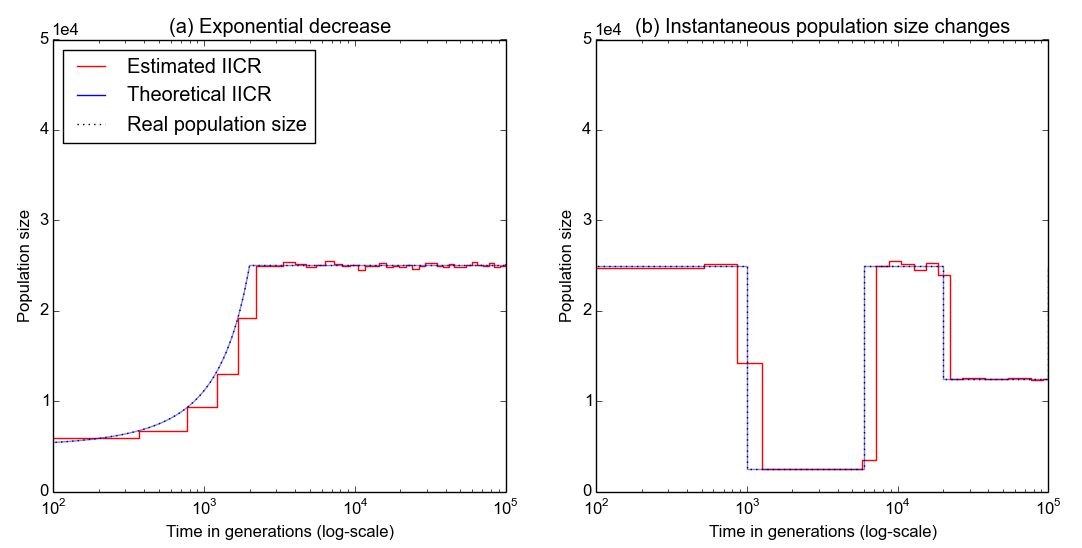}
  \caption{Inferred population size changes for populations without structure. For both panels the $x$-axis represents time in generations, whereas the $y$-axis represents population size in units of $10^4$ haploid genomes. Panel (a) represents a non structured population that experienced an exponential decrease from a previously constant size ancestral population. The solid blue line was obtained using the full theoretical $T_2$ distribution and equation \ref{Eq:lambdapsc}.  We refer to it as the theoretical IICR. The dashed line represents the total number of haploid genomes as a function of time and corresponds to the simulated demographic history. It is indistinguishable from the blue line. The stepwise red solid curve was obtained using the simulated $T_2$ values and equation \ref{Eq:IICRti}. It represents the estimated IICR. We used a limited number of time windows, but precision could easily be increased. Panel (b) shows a history of stepwise population size changes. The color codes are identical to 
panel (a). The ms-commands used for simulating the data were \textit{ms 2 100 -T -L -G -16.094 -eG 0.1 0.0 -p 8} for panel (a) and \textit{ms 2 100 -T -L -eN 0.01 0.1 -eN 0.06 1 -eN 0.2 0.5 -eN 1 1 -eN 2 2 -p 8} for panel (b).}
  \label{Fig.real_pop_size_change}
\end{figure}

Figure \ref{Fig.real_pop_size_change} shows the results for non structured populations that were subjected to various histories of population size change. The left-hand panel shows a population that experienced an exponential decrease from a previously constant size ancestral population. As expected, the  blue solid line obtained using the full theoretical $T_2$ distribution is identical to the simulated history of population size changes (\textit{i.e.} the \textit{real} population size changes). The stepwise red solid line represents the empirical IICR. The number of $t_i$ values or steps can be changed depending on the precision that one wishes to reach and the total number of $T_2$ values. We chose values similar to those typically used in recent genomic studies for comparison \citep{zhao2013panda,zhan2013psmcperegrine,zhou2014snubgenome} but a much greater precision can be achieved under our framework. The right-hand panel shows similar results but 
for a population that went through various stepwise population size changes. This shows the remarkable match between the theoretical and empirical IICR curves and the simulated history. When a population is not structured the IICR will exactly match the real history in terms of population size changes.

\begin{figure}[h]
  \centering
  \includegraphics[width=\textwidth]{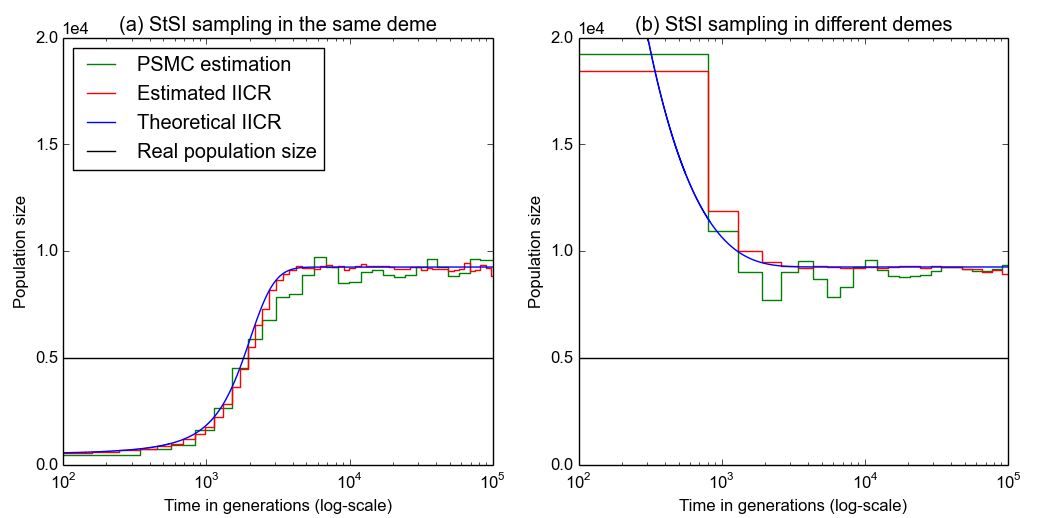} 
  \caption{Inferred population size changes under population structure and two sampling schemes.  This figure shows the predicted population size changes that will be inferred for an n-island model under the assumption that populations are not structured. We simulated a n-island model with $n=10$ and $M=1$ and computed the theoretical IICR using equation \ref{Eq:lambdapsc}, and the estimated IICR using the simulated $T_2$ values and equation \ref{Eq:IICRti}. The color codes are identical to figure \ref{Fig.real_pop_size_change}. In addition to these curves we also represent the history inferred by the PSMC (green solid line) and show that it is predicted by the IICR. Panel (a) shows the results when the two haploid genomes are sampled in the same deme. In panel (b) they come from different demes. For both panels the $x$-axis represents time in generations, whereas the $y$-axis represents real or inferred population size in units of $10^4$ haploid genomes. The constant size of the metapopulation at $y=0.5$ 
corresponds to $5,000$ haploid genomes or $10$ islands of size $500$.}
  \label{Fig.structure_sampling_effects}
\end{figure}

Figure \ref{Fig.structure_sampling_effects} is similar to Figure \ref{Fig.real_pop_size_change} but with structured populations: we sampled two haploid genomes under the StSI or n-island model, with $n=10$ and $M=1$. We used \textit{ms} to simulate both $T_2$ values and DNA sequences. We then computed the empirical IICR from the $T_2$  values, and did a PSMC analysis using the corresponding DNA sequences. Panel (a) shows the results when the genomes were sampled in the same deme (a single diploid individual) whereas panel (b) shows the results when the two haploid genomes were sampled in different demes. These figures show again that the empirical and theoretical IICR distributions match each other. Moreover they predict the population size change history inferred by the PSMC. This suggests that the PSMC does not infer a population size change but the IICR and estimates it rather well.  Finally, the IICR and the PSMC identify a (spurious) population decrease or increase depending on the sampling scheme even though the total number of haploid genomes was constant (horizontal dashed line representing the \textit{real population size}). These results are in agreement with several studies showing that different sampling strategies applied to the same set of populations may lead to infer quite distinct demographic histories \citep{chikhi2010confounding,heller2013confounding} even though they used different methods. Whereas the effect described by \cite{heller2013confounding} was observed using the Bayesian Skyline Plot method \citep{Drummond01052005}, Chikhi et al. \citep{chikhi2010confounding} used the msvar approach of Beaumont \citep{beaumont1999detecting}.

\begin{figure}[h]
  \centering
  \includegraphics[width=\textwidth]{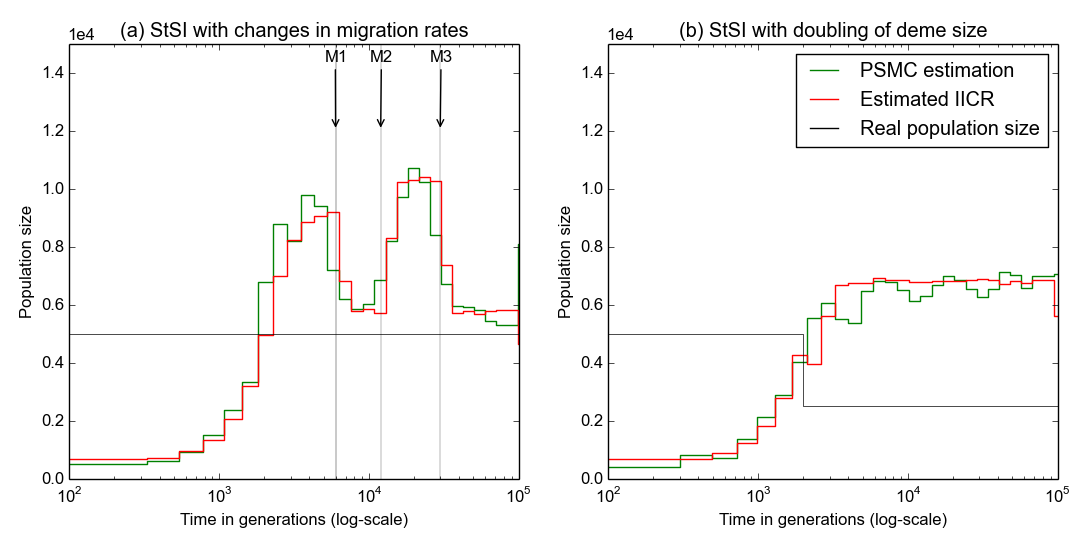}
  \caption{Inferred population size changes under population structure with changes in migration rates or deme size. Data where simulated under a StSI model with $n=10$. Color codes are identical to figure \ref{Fig.structure_sampling_effects}. The $x$-axis represents time in generations, whereas the $y$-axis represents real or inferred population size in units of $10^4$ haploid genomes. In panel (a) the population size was constant size with each deme having a size $N=500$ haploid genomes but three changes in migration rate occured at $T3=30,000$, $T2=12,000$, and $T1=6,000$ generations in the past. Before $T3$ the migration rate was $M3=5$. At $T3$ it changed to $M2=0.8$ and remained constant until $T2$, and then changed to $M1=5$ at $T1$. After that it remained at $M=1$ until the present. In panel (b) the all the demes doubled in size from $500$ to $1,000$ at $T=2,000$ generations ago. In both cases  the estimated IICR predicts the the PSMC estimation. However, interpreting the IICR as a population size 
would be misleading.The ms commands used to produce the coalescence times or DNA sequences are the following \textit{ms 2 100 -t 600 -r 120 30000000 -I 10 2 0 0 0 0 0 0 0 0 0 1 -eM 3 5 -eM 6 0.8 -eM 15 5 -p 8} for panel (a) and \textit{ms 2 100 -t 600 -r 120 30000000 -I 10 2 0 0 0 0 0 0 0 0 0 1 -eN 1 0.5 -p 8} for panel (b).}
  \label{Fig.structure_confounding_effects}
\end{figure}

While Figures \ref{Fig.real_pop_size_change} and \ref{Fig.structure_sampling_effects} illustrate and 
validate the theory developed in previous sections using two models (the StSI and PSC) for which the $T_2$ distribution is known, our approach to estimate the IICR is still valid when we have values of $T_2$ but the distribution is not known. This can happen for models that can be simulated but for which no analytical results exist 
(Figure \ref{Fig.structure_confounding_effects}). In panel (a) of Figure  \ref{Fig.structure_confounding_effects}, we considered a StSI model with $n=10$ demes where the total population size remained constant (each deme had a size of $N=1000$ haploid genomes or $N/2=500$ diploids) but migration rates changed at three different moments in the last $30,000$ generations, as indicated by the vertical arrows. This scenario mimics a set of populations whose connectivity is changing due to fragmentation or reconnection of habitat either due to climatic or anthropogenic effects. The demographic history reconstructed by the PSMC matches again the history predicted by the empirical IICR, but it is strikingly different from the actual size of the metapopulation (horizontal line). Whereas the total population size was constant throughout, the reconstructed history suggests that the population expanded and contracted on at least two occasions. A more serious issue arises from the fact that the population 
size changes inferred by the PSMC do not appear to match the times at which the migration rates changed, at least at the level of precision provided by the PSMC. For instance, the last change in migration rate, M1, occurred $6,000$ generations in the past. Still, the PSMC infers a population expansion and contraction after that event. Panel (b) corresponds to a scenario in which the size of all demes doubled $2,000$ generations before the present. Here the striking result comes from the fact that whereas the population size doubled (black broken line) the IICR and PSMC would suggest a continuous population decrease over a very long period, whose timing has again little to do with the actual history of the population. The population size change is thus missed by the PSMC. Altogether this figure suggests that changes in migration patterns or changes in deme size may be misinterpreted by population genetics methods that ignore population structure, and that there is a need for methods able to identify 
population structure from population size change (see \cite{peter2010distinguishing,Mazet201546}).

\subsection*{A tentative re-interpretation of human past demography: on the importance of being structured }

In their study \cite{li2011inference} applied the PSMC to genomic data obtained from humans and inferred a history of population size changes. As demonstrated above, what the PSMC estimates is the IICR which does not necessarily correspond to real population size changes, but may also arise from a model with changes in migration rates. To illustrate this we applied our approach to identify an island model with constant population size reproducing closely the IICR obtained  \cite{li2011inference}. For simplicity we arbitrarily assumed that the number of islands was $n=10$, and that there were three changes in migration rates as this is the minimum number of changes required to obtain an IICR curve with two humps. We propose a history in which migration rates ($M_i$, $i={1,2,3,4}$) changed at three moments ($T^i$, $i={1,2,3}$), and where $M_1$ corresponds to the number of migrants exchanged between demes each generation during the period between the present and $T^1$. More specifically, we found a change in migration rates (from $M_4=0.85$ to $M_3=0.55$) around $T^3=2.55$ million years (MY) ago, then a major increase (from $M_3=0.55$ to $M_2=4$) around $T^2=0.9-1.0$ MY and finally a major decrease (from $M_2=4$ to $M_1=0.55$) around $T^1=0.23-0.25$ MY ago. 
In other words our results would suggest changes in connectivity at the start of the Lower Pleistocene (dated at 2.58 MY), which corresponds to the emergence of the genus \textit{Homo}. The most striking change corresponds to major increase in connectivity just before the transition between the Lower and Middle Pleistocene (dated at 0.78 MY). We find that the Middle Pleistocene is characterized by high and sustained gene flow. This is a transition period during which \textit{Homo erectus} will give rise to various \textit{Homo} species including \textit{H. heidelbergensis}. Finally, connectivity  abruptly decreases at 210-230 KY ago just before the earliest remains of anatomically modern humans \textit{Homo sapiens} at \textit{ca.} 200 KY.

\begin{figure}[h]
  \centering
  \includegraphics[width=\textwidth]{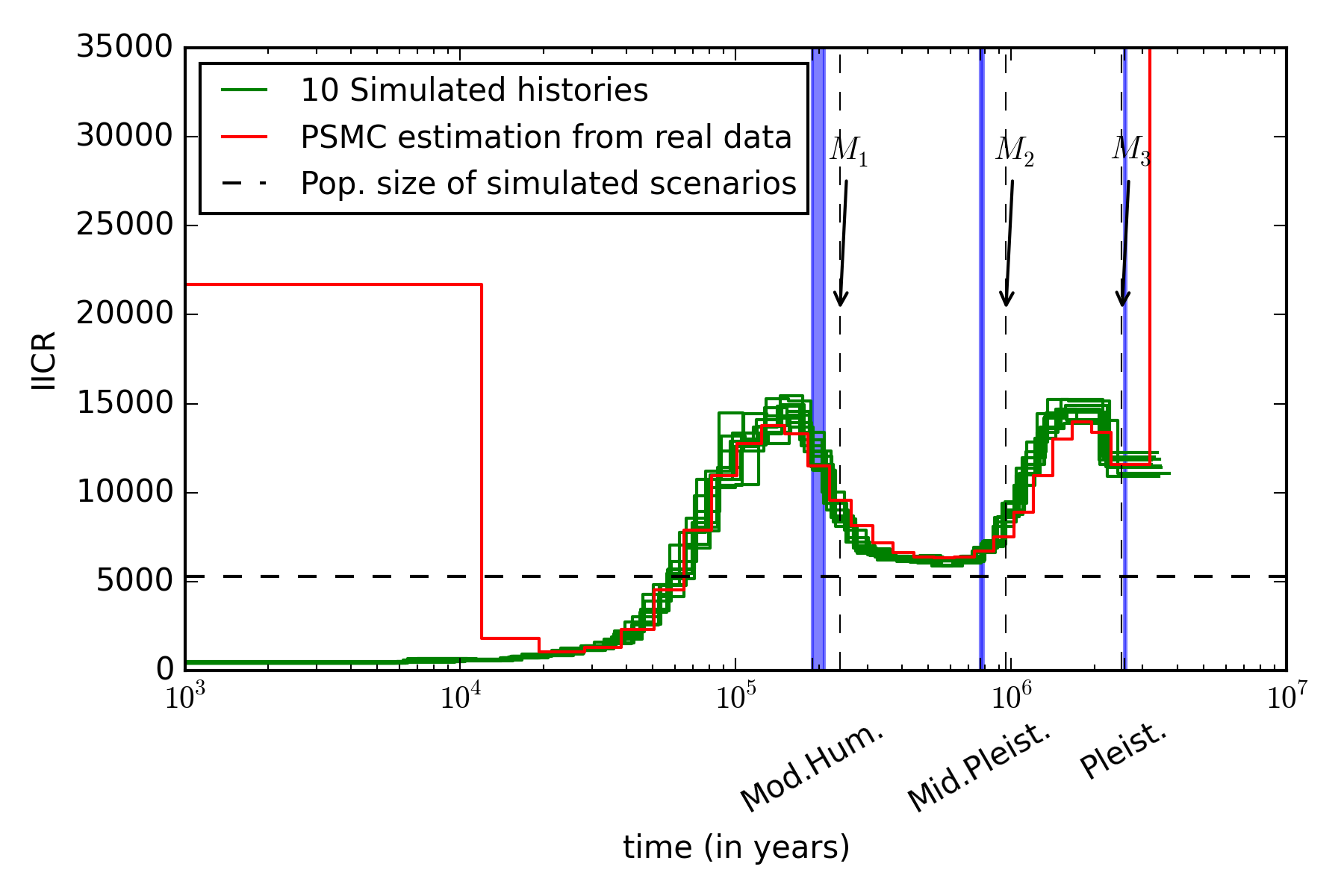}
  \caption{Human history with changes in migration rates. This figure shows, in red, the  history of population size changes inferred by Li and Durbin from the complete diploid genome sequences of a Chinese male (YH) \citep{wang2008diploid}. The $10$ green curves correspond to the IICR of ten independent replicates of the same demographic history involving three changes in migration rates. The $x$-axis represents time in years in a log scale, whereas the $y$-axis represents real or inferred population size in units of haploid genomes. The times at which these changes occur are represented by the vertical arrows at $2.55$ MY ago, $0.95$ MY ago and $0.24$ MY ago. The blue shaded areas correspond to (i) the beginning of the Pleistocene (Pleist.) at $2.57$-$2.60$ MY ago, (ii) the beginning of the Middle Pleistocene (Mid. Pleist.) at $0.77$-$0.79$ MY ago, and (iii)  the oldest known fossils of anatomically modern humans (AMH), at $195$-$198$ KY ago (\textit{i.e.} AMH must have been slightly anterior to 200 KY). Note that the PSMC is not expected to give reliable estimates of recent population sizes (\textit{i.e} less than $10$ KY), and we therefore do not present simulations with a recent demographic change due to the Neolithic transition. We focused here on the time regions for which the PSMC is expected to be reliable. However, when we do simulate a recent increase in deme size we can reproduce the increase observed in the red curve.
  The genomic data for the ten independent replicates of the scenario with three migration rate changes were simulated with ms under a StSI model with $n=10$ demes. Following Li and Durbin we assumed that the mutation rate was $\mu=2.5 10^{-8}$ and that generation time was $25$ years. We also kept the ratio between mutation and recombination rates. Each deme had a haploid size of 530 and the total number of haploid genomes was thus constant and equal to 5300. The ms command used to produce the coalescence times or DNA sequences are the following \textit{ms 2 100 -t 1590 -r 318 30000000 -I 10 2 0 0 0 0 0 0 0 0 0 0.55 -eM 4.5 4 -eM 18.0 0.55 -eM 47.5 0.85 -p 8}.}
  \label{Fig.human_hist}
\end{figure}

\section*{Discussion}

\subsection*{The IICR and the PSMC}

In this study we have shown that it is always possible to find a demographic history involving \textit{only} population size changes that perfectly explains any distribution of coalescence times $T_2$, even when this distribution was actually generated by a model in which there was no population size change. To illustrate this we first focused on a simple StSI model for which the $pdf$ of $T_2$ can be derived, and obtained an analytic formula of the fictitious population size change history, named IICR (inverse instantaneous coalescence rate), as a function of the number of islands and the migration rate of the model.
We also showed that the IICR actually exists for any (neutral) model, even if we cannot derive its theoretical $pdf$, and proposed an empirical method to estimate this IICR from any observed distribution of $T_2$ values.
Using the distribution of $T_2$ values generated by \textit{ms} for several StSI models, we showed that the empirical IICR function estimated by this method was similar to the theoretical IICR obtained previously.We then simulated $T_2$ values under complex models involving changes in migration rates or in deme size, and obtained empirical IICR in these cases as well. This suggests that, at least for a sample of size 2, even an infinite amount of genetic data from independent loci alone could not distinguish extremely different models. The model of population structure in which we varied migration rates at various times is profoundly different from the model of expansions and contractions represented in Figure \ref{Fig.structure_confounding_effects}. Also, the history of population size changes reconstructed by methods ignoring structure would suggest that four demographic changes occured, two expansions and two contractions, whereas only three changes of the migration rate were actually simulated.

The theory presented here is simple, but it allowed us to predict the history reconstructed by any method ignoring population structure and to re-interpret the parameters estimated by any such  method. What most population genetics methods will try to estimate is the IICR. This IICR corresponds to the population size only in cases where the population is non structured. In the case of more complex demographic histories, interpreting the IICR as a population size or a ratio of population sizes can be very misleading. To clarify the difference between the IICR and an effective population size we can consider the following rationale. If a structured population could be summarized by a single $N_e$ then a change in gene flow should be matched by a simultaneous change in $N_e$. In that case, changes in $N_e$ would be misleading (since the size would not change) but their timing might still be meaningful. For instance a ``hump'' inferred using diCal, the PSMC or the MSMC, among others could be easily translated into a change in gene flow patterns. In such a case, we could re-interpret the changes in $N_e$ by saying, for each hump, that gene flow decreased and then increased again. What the IICR shows is that it is not that simple. The fact that a structured model can only be summarized by a \textit{trajectory} of spurious population sizes means that the timing of changes in migration rates will interact in a complex manner hence generating IICR profiles that may be only loosely related with populational events. This can be seen in Figure 5 and 6.

These results do not invalidate the use of panmictic models for the reconstruction of population history as long as population structure can indeed be neglected, but it certainly stresses the need for caution in the interpretation of this history. When Li and Durbin published their landmark study in 2011 they showed for the first time that it was possible to reconstruct the demographic history of a population by using the genome of a single diploid individual \citep{li2011inference}. It was a remarkable feat based on the SMC model introduced by \cite{mcvean2005coalrecomb}. Its application to various species \citep{prado2013psmcgreatapes,zhou2014snubgenome,groenen2012pig,zhao2013panda,green2014crocodile,zhan2013psmcperegrine,hung2014pigeon} has been revolutionary and led to the development of new methods \citep{Sheehan22042013, schiffels2013msmc}. However, the increasing number of studies pointing at the effect of population structure \citep{leblois2006genetics,nielsen2009statistical,chikhi2010confounding,heller2013confounding,paz2013demographic} or changes in population structure \citep{wakeley1999nonequilibrium,stadler2009impact,broquet2010genetic,heller2013confounding,paz2013demographic} in generating spurious changes in population size suggested that new models should be analysed that can incorporate population structure \citep{goldstein2002human,harding2004structured}. For instance,  \cite{Mazet201546} have recently shown that genomic data from a single diploid individual can be used to distinguish an n-island model from a model with a single population size change. This study represents an interesting alternative since it should be possible to determine whether a model of population structure is more likely than a model of population size change to explain a particular data set. The approach of \cite{Mazet201546} is however limited to a very simple model of population size change. 
Demographic models inferred by several recent methods \citep{li2011inference,schiffels2013msmc,Sheehan22042013} are not limited to one population size change. They are thus more realistic, and, as we have shown here this comes at a certain price. Since they allow for several tens of population size changes, they mimic more precisely the genomic patterns arising from structured models. Therefore, they reconstruct a demographic history that can optimally explain any particular pattern of genomic variation only in terms of population size changes. As we have shown here, and until we can separate models (see below) this casts doubts on any history reconstructed from genomic data by the above-mentioned approaches. Indeed, if any pattern of (neutral) genomic variation can be interpreted efficiently in terms of population size changes, then how can we identify the cases where the observed genomic data were not generated by population size changes?

It is worth noting that \cite{li2011inference} acknowledged that one should be cautious when interpreting the changes inferred by their method. For instance, they showed (see their Supplementary Materials) that when one population of constant size $N$ splits in two half sized populations that later merge again, their method will identify a change of $N$ even though $N$ actually never changed. Still, their method is implicitly or explicitly used and interpreted in terms of population size changes, including by themselves. There are therefore several issues that need to be addressed. One issue is to determine whether it is possible to separate models of population size change from models of population structure (\citep{Mazet201546}, see below). When population structure can be ignored, our results actually contribute to the validation of the PSMC. We found that the PSMC performed impressively well and generally reconstructed the IICR with amazing precision. It is therefore at this stage one of the best methods \citep{Sheehan22042013,schiffels2013msmc} published so far and remains a landmark in population genetics inference.

\subsection*{The IICR: towards a critical interpretation of effective population sizes}

The concept of effective size is central to population genetics. It allows population geneticists to replace complex real-world populations by equivalent and simpler Wright-Fisher populations \textit{that would have the same ‘‘rate of genetic drift.’’} \citep{wakeley2009extensions}. The concept is however far from trivial and it is not always clear what authors mean when they mention the $N_e$ of a particular species or population, as rightly noted by \cite{sjodin2005meaning} among others. Several $N_e$s have been defined depending on the property of interest (inbreeding, variance in allele frequency over time, etc.) and its relationship to genetic drift \citep{wakeley2009extensions}. This is a complex issue which we do not aim at reviewing or discussing in detail here.
 
 The IICR is related to the coalescent $N_e$ \citep{sjodin2005meaning,wakeley2009extensions} but it is explicitly variable with time. 
Given that most species are likely to be spatially structured, interpreting the IICR as a simple (coalescent) effective size may generate serious misinterpretations.

Our results suggest that there is a trajectory of population sizes, the IICR, which fully explains complex models without loss of information. Whether this trajectory can indeed and under some circumstances be appropriately summarized by one effective population size is still to be determined. For instance, for $M=500$ and $M=100$ the corresponding trajectories are characterized by population size changes that are recent and abrupt. Under such ``strong migration scenarios'' it is probably acceptable to consider the period during which the population was stationary as most significant to generate patterns of genetic diversity \citep{wakeley1999nonequilibrium,charlesworth2003structure,wakeley2009extensions} . However, even for such cases of low genetic differentiation ($F_{ST}$ $\approx$ $1/2001=0.0005$ and $F_{ST}$ $\approx$ $1/401=0.0025$, respectively), the spurious population size drop could perhaps be detected with genomic information. For $M=100$ the population size decrease starts between $t=0.05$ and $t=0.10$, which for $N=100$ to $N=1000$ could correspond to values between $5$ to $100$ generations ago, respectively. In other words, a StSI model may actually behave differently from a WF model even under some ``strong migration'' conditions. The approximation will therefore be valid for some questions and data sets, and invalid for others \citep{charlesworth2003structure,wakeley2009extensions} .

\subsection*{The IICR and the complex history of species: towards a critical re-evaluation of population genetics inference}

The PSMC has now been applied to many species, generating curves that are very similar to those represented in Figure \ref{Fig.structure_confounding_effects}. In the specific case represented in panel (a) our results suggested that the expansions and contractions detected by the PSMC were not correlated in a simple manner to the changes in gene flow or deme size. This lack of simple correlation is likely the result of two factors. The first is that, as we showed, a structured population cannot always be summarized by a single number. The second factor is that the PSMC uses a discretized distribution of time which may lead to missing abrupt changes such as those simulated here. For real data sets where changes in migration rates or in population size may be smoother, this may not be so problematic. Our analysis of the human data, however, led us to propose a scenario that is profoundly different from current interpretations of genetic and genomic data. Assuming a simple model of population structure we identified periods of change in gene flow which correspond to major transitions in the human recent evolutionary history, including the emergence of the \textit{Homo} genus at 2.55 MY strikingly close to the start of the Pleistocene (a 2.58 MY), the beginning of the Middle Pleistocene and finally the emergence of anatomically modern humans. Given that humans are likely to have been subjected to a complex history of spatial expansions and contractions and changes in the levels of gene flow \citep{harpending2000genetic,goldstein2002human,harding2004structured}, our results suggest that a re-interpretation of panmictic models may be needed and possible.
Our results are at odds with a history of population crashes and increases depicted in many population genetic studies, but it is in phase with fossil data and provides a more realistic framework. We thus wish to call for a critical reappraisal of what can be inferred from genetic or genomic data. The histories inferred by methods ignoring structure represent a first approximation but they are unlikely to provide us with the information we need to better understand the recent evolutionary history of humans or other species. It is difficult to imagine a panmictic population whose size has changed over the last few million years. This does not minimize the achievement of the \cite{li2011inference} study, but it does question how genetic data are sometimes interpreted.

\subsection*{Perspectives}

We focused throughout this study on $T_2$, but it would be important to  determine whether, for structured models, the IICR estimated from the distribution of $T_k$ varies significantly with $k$. If that were the case, that would suggest that it is possible to separate structured models from population size change models with the distributions of $T_k$ for various $k$ values. The reason for this is that population size change models should generate identical IICR for all $T_k$ distributions, since they should all correspond to the same (real) history of population size change. To our knowledge the distribution of $T_k$ for $k>2$ has not yet been derived for the StSI or other classical structured models. 

One possible and relatively simple solution to this question is to simulate genetic data under a structured model of interest and then compare the simulated $T_k$ distributions under that model and the $T_k$ distributions of the corresponding model of population size change identified using the $T_2$ distribution. Preliminary simulations suggest that the $T_k$ distributions produce different IICRs, at least for some models of population structure. For instance, we predict that the analysis of human genomic data with the PSMC and with the MSMC should produce different curves under a model of population structure but identical ones for a model of population size change. This prediction can be tested by comparing the PSMC and MSMC curves of \cite{li2011inference} and \cite{schiffels2013msmc}, respectively. Visual inspection of the corresponding figures suggests indeed that they are different, and therefore that our model of population structure is a valid alternative. However, we stress that an independent study is required. Indeed, the history reconstructed by these methods with real data is not very precise and the two curves are not easily comparable because they are expected to provide poor estimates at different moments. Any difference between the two analyses should thus be evaluated and validated with simulations.

There is another point which we would like to mention. One underlying assumption of our study is that the coalescent represents a reasonable model for the genealogy of the genes sampled. Given that the coalescent is an approximation of the true gene genealogy, and that there are species for which the coalescent may not be the most appropriate model \citep{wakeley2009extensions} we should insist that our results can, at this stage, only be considered for coalescent-like genealogies. The development of similar approaches for other genealogical models would definitely be a very interesting avenue of research.

\section*{Acknowledgements}
We are grateful to A. Lambert and G. Achaz for constructive comments on earlier versions of this work. We are grateful to the genotoul bioinformatics platform Toulouse Midi-Pyrenees for providing computing and storage resources and to the LIA BEEG-B (Laboratoire International Associé - Bioinformatics, Ecology, Evolution, Genomics and Behaviour) (CNRS) for facilitating travel and collaboration between Toulouse and Lisbon. This work was partly performed using HPC resources from CALMIP (Grant 2012 - projects 43 and 44) from Toulouse, France. This study was partly funded by the Fundação para a Ciência e Tecnologia (ref. PTDC/BIA- BIC/4476/2012), the Projets Exploratoires Pluridisciplinaires (PEPS 2012 Bio-Maths-Info) project, the LABEX entitled TULIP (ANR-10-LABX-41) as well as the Pôle de Recherche et d'Enseignement Supérieur (PRES) and the Région Midi-Pyrénées, France.

\bibliography{Bibliography}

\end{document}